\documentclass{aastex}
\slugcomment{submitted to ApJ Letters}
\shorttitle{Mass Profile of CL~0024+1654 from Lensing}
\shortauthors{Shapiro \& Iliev}
\begin{document}
\title{On the Mass Profile of Galaxy Cluster CL~0024+1654
Inferred from Strong Lensing}
\author{Paul R. Shapiro}
\affil{Department of Astronomy, University of Texas, Austin, 78712}
\email{shapiro@astro.as.utexas.edu}
\and
\author{Ilian T. Iliev }
\affil{Department of Physics, University of Texas, Austin, 78712}
\email{iliev@astro.as.utexas.edu}

\begin{abstract}
Observations of a flat density profile in the cores of
 dark-matter-dominated halos on the
two extremes of mass for virialized objects in the universe, dwarf galaxies
and galaxy clusters, present a serious challenge to the current standard 
theory of structure formation involving Cold Dark Matter (CDM). By contrast, 
N-body simulations of halo formation in the latter indicate density profiles
which are singular and steeply rising towards the center. A flat-density core
on the cluster scale is indicated by gravitational lensing observations, most
significantly by the strong-lensing measurements of CL~0024+1654 by the
Hubble Space Telescope. A 
recent re-analysis of this cluster has suggested that a uniform-density core
is not demanded by the data, thereby eliminating a significant piece of the
conflict between the observations and the CDM theoretical predictions. We
show here, however, that the singular mass profile which that analysis
reports as consistent with the lensing measurements of CL~0024+1654
implies a velocity dispersion which is much higher than the measured value for
this cluster.  
\end{abstract}


\section{Introduction}
\special{}
There has been a lot of recent controversy concerning the density profiles of
the dark matter halos of virialized cosmological structures, from dwarf galaxy
to galaxy cluster scale. N-body simulations of the formation of ``virialized''
dark matter halos associated with galaxies and clusters in a Cold Dark Matter 
 model were found to be well-fit by a simple, universal form for the 
variation of mass density $\rho$ with radial distance $r$ from the center of
mass, given by
\begin{equation}
\label{NFW}
\rho_{\rm NFW}(r)={\rho_S\over(r/r_S)(1+r/r_S)^2},
\end{equation}
where $r_S$ is some characteristic radius which separates the two 
asymptotic power-law slopes, $\rho\propto r^{-1}$ at $r<<r_S$ and 
$\rho\propto r^{-3}$ at $r>>r_S$, and $\rho_S$ is a characteristic
density which reflects the mean density of the universe at the epoch
of halo formation (Navarro, Frenk, \& White 1997; ``NFW''). More 
recent N-body simulations of higher resolution obtain halo profiles 
which agree with the NFW profile at large radii but have an even 
steeper inner slope, $\rho\propto r^{-1.5}$ (Moore { et al.} 1999). 

This prediction by the standard CDM model of singular density profiles
for cosmological halos is apparently in conflict with the observed 
mass distributions inside dark-matter-dominated halos on two extremes 
of the halo mass function, dwarf galaxies and galaxy clusters.
As a result, these observations and their interpretation have
recently come under intense scrutiny. In work going back to 
Flores \& Primack (1994), Moore (1994), Burkert (1995) and
more recently in Kravtsov { et al.} (1998), Moore { et al.} (1999), and 
McGaugh \& de Blok (1999), these predictions of singular halos were
 found to be in conflict with the halo density profiles derived 
from the observed rotation curves of dark-matter-dominated dwarf galaxies. 
The latter were found, instead, to be better fit by density profiles with 
flat cores.
Recently, the universality of this observational requirement that 
dark-matter-dominated galactic halos possess a uniform-density core
has been challenged on the grounds that the rotation curves are not
generally resolved well enough in the centers to distinguish the
 slowly rising rotation curve which results from a mass profile with 
a flat core from the more rapid rise which results from a cuspy profile \cite{BRDB}.
Initially, this challenge appeared to apply primarily to LSB galaxies, leaving
the case for uniform-density cores in dwarf galaxies still strong.
For example, the well-resolved rotation curves of the nearby dwarf galaxies
DDO 154 and NGC 3109 still demanded a flat density core \cite{BRDB}. 
Very recently, however, this challenge has been extended to
the interpretation of dwarf galaxy rotation curves, as well 
(van den Bosch \& Swaters 2000). While the latter are still generally 
better fit by halo profiles with a flat core, it can no longer be stated with
much confidence that the observations {\it demand} such profiles. Until 
better 
resolution data become available, profiles with  $\rho\propto r^{-\alpha}$
near the center, with $0\leq\alpha\leq1$, all seem to yield reasonable fits
to the data. Hence, while it appears that a significant 
conflict remains between the CDM halos predicted by the highest resolution
N-body simulations to date (which imply $\rho\propto r^{-1.5}$ at the 
center) and the dwarf galaxy rotation curves, the current observations are
unable to discriminate effectively between halos with inner profiles as
cuspy as the NFW profile and those with a uniform-density core, instead.

On the galaxy cluster scale, the case for a uniform-density core has been made
most convincingly using observations of strong gravitational lensing, where 
the images of 
background galaxies are distorted by the cluster mass to form arcs and multiple
images. To date, the most spectacular example of such observation is a Hubble
Space Telescope image of multiple arcs produced by the cluster CL~0024+1654 
at $z=0.39$. The relaxed structure 
of CL~0024+1654, the absence of a single, central, dominant cluster galaxy, 
and the presence of an easily-identified multiply-imaged background galaxy
make this cluster a particularly good candidate for a determination of the
halo mass profile from lensing analysis. According to Tyson,  Kochanski, \& 
Dell'Antonio (1998), the observations of this cluster require a halo mass
profile with a uniform-density core, in strong conflict with the predicted
cuspy profile of NFW.

These conflicts between the CDM N-body results and the observations of dwarf
galaxy and cluster halo profiles have stimulated a vigorous re-examination 
of the 
theoretical underpinnings of the CDM model, including a number of suggestions 
for a variation of the microscopic properties of CDM which might serve to
produce halos with uniform-density cores while retaining the more successful 
aspects of the original CDM model. These include suggestions that the dark 
matter is nongravitationally self-interacting \cite{SS}, warm (e.g. 
Sommer-Larsen \& Dolgov 2000; Col\'\i n, Avila-Reese, \& Valenzuela 2000; 
Hannestad \& Scherrer 2000), fluid \cite{P}, decaying \cite{C}, repulsive 
\cite{G}, fuzzy \cite{HBG}, and annihilating \cite{KKT}. In view of the 
importance of the dwarf galaxy and cluster halo profile observations in 
constraining the theoretical models, it is somewhat disappointing that 
the conclusions based upon dwarf galaxy rotation curves are currently so 
ambiguous with regard to the question of the uniform-density core. This 
makes the conclusion of Tyson { et al.} (1998) regarding the core in 
CL~0024+1654 all the more critical.  

Recently, a new study of this cluster by Broadhurst { et al.} (2000)
has reached a conclusion opposite to that of Tyson { et al.} (1998) regarding
the consistency of the observed mass profile with the density profiles predicted 
by the N-body simulations of cluster formation in the standard CDM model.
They find that the NFW mass profile is consistent with the lensing data.
If this is correct, then the case against the standard CDM model is 
significantly weakened. The purpose of this letter is to point out that the 
fit by Broadhurst { et al.} (2000) of the cluster lensing data with a mass
distribution which follows the NFW profile implies a cluster velocity 
dispersion which is much larger then the value measured for this cluster 
by Dressler et al. (1999) of $\sigma_V= 1150\,{\rm km\,s^{-1}}$.

\section{Observational Results for the Cluster Mass Profile}

The projected dark matter density profile which Tyson { et al.} (1998)
found by modeling the lensing data for CL~0024+1654
within the arcs at radius $r_{\rm arcs}\sim100\,h^{-1}{\rm kpc}$
is well fit by 
\begin{equation}
\label{lens_fit}
\displaystyle{\Sigma(y)=\frac{K(1+\eta y^2)}{(1+y^2)^{2-\eta}}},
\end{equation}
where $y=r/r_{\rm core}$, $K=7900\pm100\,h\,{\rm M_\odot\,pc^{-2}}$, 
$r_{\rm core}=35\pm3\,h^{-1}\,{\rm kpc}$, $\eta=0.57\pm0.02$, and
$h$ is the Hubble constant in units of $100\, {\rm km\,s^{-1}Mpc^{-1}}$.  
Additionally, Tyson { et al.} (1998) rule out at a great confidence level
the possibility of a good fit of the observed mass distribution 
by the NFW profile.

A recent paper by Broadhurst { et al.} (2000) obtained, 
instead, a good fit to the lensing data with a total mass 
distribution given by the NFW profile in equation~(\ref{NFW}),
with $r_S\approx 400\,h^{-1}\,{\rm kpc}$ and
$\delta_c\approx8000$, where $\delta_c\equiv\rho_S/\rho_{\rm crit}(z)$,
and $\rho_{\rm crit}(z)\equiv 3H^2/(8\pi G)$, the critical density of the
universe at the cluster redshift. 
This $\delta_c$ is 
directly related to the NFW concentration parameter $c\equiv r_{\rm 200}/r_S$,
where $ r_{\rm 200}$ is the radius within which the average density is 200
times this critical density, according to
\begin{equation}
\label{delta_c}
\displaystyle{\delta_c=\frac{200}{3}\frac{c^3}{[\ln(1+c)-c/(1+c)]}},
\end{equation}
which yields $c\approx5$, and $r_{200}\approx 2\,h^{-1}$Mpc. 
From this they concluded that there is no conflict between the 
observations of this cluster and the predictions of the standard CDM model. 


\section{Consequences for the Cluster Velocity Dispersion}
\subsection{NFW profile}
These mass profiles inferred for the cluster CL~0024+1654 based upon lensing 
observations have implications for the velocity dispersion of its dark
matter and galaxies if the cluster is assumed to be in virial
equilibrium. We begin by considering the NFW profile.
The same N-body simulations of cluster formation in the
CDM universe which indicate that clusters are, indeed, in an approximate
virial equilibrium which can be described by the universal mass profile
suggested by NFW also yield information about the 1-D velocity dispersion 
$\sigma_V$ and its radial dependence. 
Over a wide range of radii, the halos obtained in N-body simulations 
are roughly isothermal
(Tormen, Bouchet, \& White 1997; Eke, Navarro, \& Frenk 1998).
For comparison with the velocity dispersion of a cluster like
CL~0024+1654 observed within some radius, it is necessary to consider 
the average velocity dispersion of the NFW halo within a sphere of the 
same radius. The average velocity dispersion 
of CL~0024+1654 was measured by Dressler et al. (1999) to be 
$\sigma_V=1150\,{\rm km\,s^{-1}}$ within a 
radius $r\sim 600\,{\rm h^{-1}kpc}\approx6r_{\rm arcs}$, based upon
107 galaxy redshifts, to an uncertainty of roughly less than 
$\pm100\,{\rm km\,s^{-1}}$. For the NFW profile 
proposed by Broadhurst { et al.} (2000) for this cluster, this radius 
corresponds to $r\approx r_{\rm 200}/3$.
It is a relatively simple matter to estimate the predicted velocity dispersion
$\sigma_{\rm V,NFW}$ for a given NFW density profile in terms of the 
circular velocity profile $v_{\rm c,NFW}$ of the same halo, as follows.

The circular velocity profile of the NFW halo is given by
\begin{equation}
\label{rot_curve}
\displaystyle{v_{\rm c,NFW}^2(r)\equiv\frac{GM(\leq r)}r
	= 4\pi G\rho_Sr_S^2\frac{[\ln({x}+1)-{x}/{(x+1)}]}{x}}, 
\end{equation}
where $M(\leq r)$ is the mass enclosed by radius $r$, and
$x\equiv r/r_S$. The maximum value of $v_{\rm c,NFW}$ is
$v_{\rm max,NFW}\approx0.465( 4\pi G\rho_Sr_S^2)^{1/2}$, which occurs at 
$x=x_{\rm max}\approx2.163$. The NFW velocity profile for the
parameters reported by Broadhurst { et al.}~(2000) is shown in 
Figure~1. We obtain $v_{\rm max,NFW}\approx 3340\,{\rm km\,s^{-1}}$ for 
this cluster. According to the detailed analysis of numerical N-body
results by Tormen, Bouchet, \& White (1997), the average 1-D velocity 
dispersion within $r\approx r_{200}/3$ for simulated clusters, which 
are well-fit by the NFW profile, is somewhat 
lower than $v_{\rm max,NFW}$, but never by more than a factor of 
$\approx1.5$. This factor of 1.5 agrees very well with the 
aperture-averaged, line-of-sight $\sigma_V$
which results from solving the Jeans equation for the variation of the 
radial $\sigma_{V,r}$ with $r$ inside the NFW profile, including the
possible effects of anisotropic velocities (Lokas \& Mamon 2000; 
Lokas, private communication).
Therefore, the NFW profile proposed for CL~0024+1654 by Broadhurst { et al.}
(2000) implies an average velocity dispersion for the cluster within
the radius $r\sim 600\,{\rm h^{-1}kpc}\approx6r_{\rm arcs}$ of
$\sigma_{\rm V,NFW}>v_{\rm max,NFW}/1.5\sim2230\, {\rm km\,s^{-1}}$, 
which is too large by a factor of $\sim 2$
to be compatible with the measured velocity dispersion quoted above
\footnote{This assumes $\rho_S\approx 0.006\,h^2\,{\rm M_\odot/pc^3}$,
the value if $\rho_{\rm crit}(z)$ is for an 
Einstein-de~Sitter universe. Our conclusion that $\sigma_{V,NFW}$ substantially
exceeds the observed $\sigma_{\rm V}$ remains unchanged if, instead, we consider
other background cosmologies. The values of $v_{\rm max,NFW}$ and 
$\sigma_{\rm V,NFW}$ for a low-density universe ($\Omega_0=0.3$)
are only slightly lower than the values quoted above, by 10\% for
the open case ($\Lambda=0$) and by 25\% for the flat 
($\Lambda\neq0$) case.}.

\subsection{Nonsingular Isothermal Sphere}
The Tyson { et al.} (1998) mass model in equation~(\ref{lens_fit}) does
not uniquely predict the velocity dispersion $\sigma_V$ implied for the 
cluster even if it is assumed to be in virial equilibrium. For one,
it is a fit {\it only } to the projected density within the arcs at 
$r_{\rm arcs}\sim100\,{\rm h^{-1}kpc}$ and, therefore, does not 
constrain the external mass profile very well. For another, even if 
we restrict our attention to radii within
the arcs and assume virial equilibrium, this is still not enough to specify 
$\sigma_V$ and its radial dependence there uniquely. As a result, it 
is not possible to compare the predicted $\sigma_V$ for the 
Tyson { et al.} (1998) mass model with the observed $\sigma_V$
as directly as we did for the NFW profile
above. However, we can at least demonstrate that the mass model is plausibly
consistent with the observed $\sigma_V$, as follows.

The projected mass density in equation~(\ref{lens_fit}) is well fit by
that obtained from the truncated, nonsingular, isothermal sphere 
(``TIS'') profile of Shapiro, Iliev, \& Raga (1999). We discuss the 
details of the fitting procedure and the results elsewhere (Iliev \& 
Shapiro 2000a,b). This TIS represents a particular solution of
the Lane-Emden equation which corresponds to the outcome of the collapse
and virialization of a top-hat density perturbation.
The size and virial temperature which result are unique functions of 
the mass and redshift of formation of the object for a given background 
universe. According to this solution, the central density $\rho_0$ is
roughly more than 500 times the density at the surface, and the core radius 
$r_0$ is about 1/30 of the total size. [Note: Our definition of the core radius 
is $r_{\rm 0,TIS}\equiv r_{\rm King}/3$, where $r_{\rm King}$ is the
``King radius'' defined in Binney \& Tremaine (1987), p. 228.] As described 
elsewhere, this solution is a convenient analytical approximation for
the halos which form from  more realistic initial conditions in the CDM
model, which reproduces many of the average structural properties of the
halos found in CDM simulations, except in the very inner profile where
the TIS model has a uniform-density core instead of a central cusp. 
As such, a fit of the TIS profile to the Tyson { et al.} (1998)
mass model provides a plausible, physically-motivated connection between
this mass model with a flat density core and the implied cluster velocity
dispersion. The projected density profile of a TIS sphere with a central density  
$\rho_0\approx0.064\,h^2\,{\rm M_\odot/pc^3}$ and a core radius 
$r_0\approx20\,h^{-1}\,$kpc provides a very close match to the 
Tyson { et al.} (1998) result for $\Sigma(r)$ discussed above.
Based on this best-fit TIS, the velocity dispersion is
\begin{equation}
\sigma_{\rm V,TIS}=(4\pi G\rho_0r_0^2)^{1/2}\approx1200\,{\rm km\,s^{-1}},
\end{equation}
in close agreement with the measured value. 

\section{Conclusion}
Two different attempts to invert the observations of strong lensing
by cluster CL~0024+1654 to solve for the mass profile of the cluster
have reached opposite conclusions regarding the presence of a 
uniform-density core versus the acceptability of a central cusp like 
that of the NFW profile. This suggests that, either there is some error 
in one or both of these analyses or else there is some degeneracy in the
inversion process which prevents a clean discrimination between these  
different solutions. While we do not claim to address the accuracy
of either of the two analyses of this cluster, we point out, instead,
that there may be additional constraints on the allowed mass models 
which can aid in distinguishing them. In particular, the measured 
cluster velocity dispersion, which also reflects the mass distribution
of the cluster, should also be required to be consistent with the
mass model derived from the inversion of the lensing data. We have 
shown that this is not the case for the cuspy NFW profile which
Broadhurst { et al.} (2000) report is a good fit to the lensing data.
That profile, if it actually corresponds to a halo formed in the 
standard CDM model, predicts much too high an average velocity dispersion  
to be consistent with the observed value. By contrast, the mass model of 
Tyson { et al.} (1998) does not have this problem.
This suggests that the lensing data for CL~0024+1654 still favor
a flattening of the density profile at small radii, in conflict with 
the prediction of cuspy halos by N-body simulations of the standard
CDM model. 

In view of the importance of this conclusion for the ongoing debate 
regarding the validity of the standard CDM model, it would be valuable 
if uncertainties in the gravitational lense models, such as that 
due to possible departures from spherical symmetry or substructure, 
were properly quantified. This uncertainty would have to be rather 
extreme, however, to reconcile the Broadhurst et al. (2000) profile
fit with the observed velocity dispersion of CL~0024+1654. The
lensing analysis leads to NFW parameters roughly by measuring the 
mass interior to the arcs, 
$M(\leq r_{\rm arcs})=4\pi\rho_S r_S^3f(x_{\rm arcs})$, where 
$f(x)=\ln(1+x)-x(1+x)^{-1}$ and $x\equiv r/r_S$, and determining 
$r_S$ by matching the logarithmic slope $\gamma$ of the density 
profile at $r_{\rm arcs}$ as derived from the projected mass 
distribution, according to $x_{\rm arcs}=-(1+\gamma)/(3+\gamma)$.
In order to adjust $v_{\rm max,NFW}$ downward relative to the value
discussed above by a factor large enough to reconcile the NFW profile
with the observed velocity dispersion while leaving the measured mass
$M(\leq r_{\rm arcs})$ unchanged, therefore, $r_S$ must be reduced by
factor of order 4, to $r_S\approx 100\,h^{-1}{\rm kpc}$. This requires
an observational uncertainty so large as to change the value 
$\gamma\approx -1.3$ reported by Broadhurst et al. (2000) into 
$\gamma=-2$, which seems very unlikely.

\acknowledgments
We are grateful to Alan Dressler for valuable discussion and to
the referee for comments which led us to improve the paper.
This work benefited from the partial support of grants 
NASA ATP NAG5-7821, NSF ASC-9504046, and Texas 
Advanced Research Program 3658-0624-1999.

\newpage
\figcaption[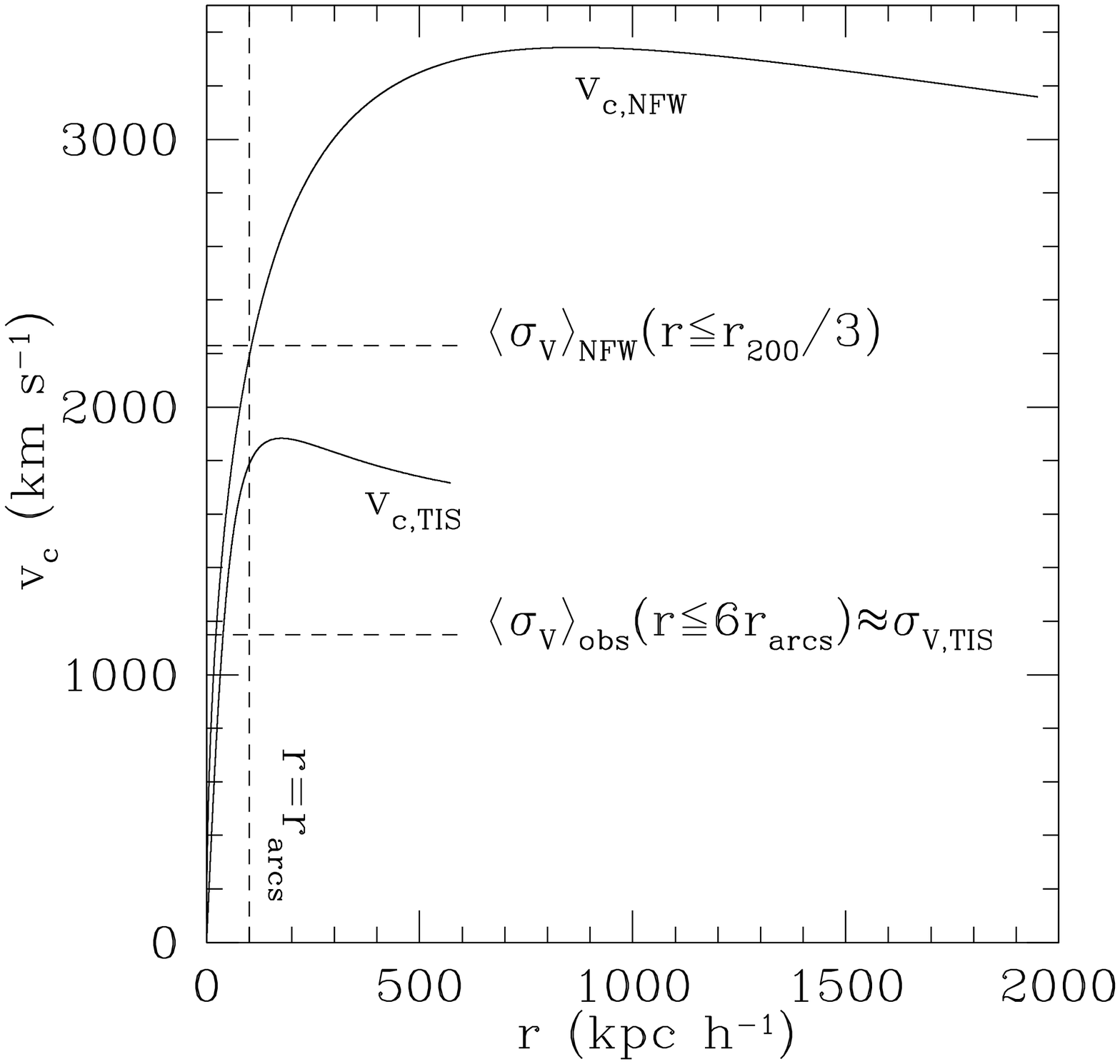]{Circular velocity profiles for cluster 
CL~0024+1654 implied by
the mass profiles inferred from the modelling of strong gravitational 
lensing data by Broadhurst { et al.} (2000) (solid curve labelled 
``$v_{\rm c,NFW}$'') and Tyson { et al.} (1998) [when the latter is fit 
by the truncated, nonsingular isothermal sphere model of Shapiro { et al.} (1999)]
(solid curve labelled ``$v_{\rm c,TIS}$''.) Horizontal dashed lines show 
the average velocity dispersions $\langle\sigma_V\rangle$, predicted for 
each mass model, as labelled, along with the value observed by 
Dressler et al. (1999). Vertical dashed line indicates the radius of the 
arcs, within which the lensing data constrains the mass profile.}
\begin{figure}
\plotone{fig1.eps}
\end{figure}
\end{document}